\documentclass[superscriptaddress,onecolumn,showkeys,preprintnumbers,amsmath,amssymb]{revtex4}
\usepackage{graphicx}
\usepackage{dcolumn}
\usepackage{bm}
\usepackage{color}

\begin{document}

\preprint{Published in \textit{Trends in Mathematics} {\bf{2}}: 1-4 (2014) (Springer)}

\title{Viral RNA replication modes: evolutionary and dynamical implications}

\author{Josep Sardany\'es}
\thanks{E-mail: josep.sardanes@upf.edu}
\affiliation{ICREA-Complex Systems Lab, Departament de Ci\`encies Experimentals i de la Salut (Universitat Pompeu Fabra), Dr. Aiguader 88, 08003 Barcelona, Spain}
\affiliation{Institut de Biologia Evolutiva (CSIC-Universitat Pompeu Fabra), Passeig Maritim de la Barceloneta 37, 08003 Barcelona, Spain.} 




\begin{abstract} 
Viruses can amplify their genomes following different replication modes (RMs) ranging from the stamping machine replication (SMR) model to the geometric replication (GR) model. Different RMs are expected to produce different evolutionary and dynamical outcomes in viral quasispecies due to differences in the mutations accumulation rate. Theoretical and computational models revealed that while SMR may provide RNA viruses with mutational robustness, GR may confer a dynamical advantage against genomes degradation. Here, recent advances in the investigation of the RM in positive-sense single-stranded RNA viruses are reviewed. Dynamical experimental quantification of Turnip mosaic virus RNA strands, together with a nonlinear mathematical model, indicated the SMR model for this pathogen. The same mathematical model for natural infections is here further analyzed, and we prove that the interior equilibrium involving coexistence of both positive and negative viral strands is globally asymptotically stable. 
\end{abstract}


\maketitle

\section{Introduction}
RNA viruses are obligate parasites infecting bacteria, fungi, plants and animals. Upon infection, RNA viruses replicate within the host cells generating a highly heterogeneous population of viral genomes named quasispecies \cite{Domingo2001}. Generally, the viral replicase copies the initially infecting positive-sense or genomic strand, producing the negative or antigenomic one. How these templates are then processed for further replication has been a subject of research, and different replication modes (RMs) have been proposed. For instance, if the produced negative template is mainly used as a template for the production of the whole progeny of genomic strands, the linear stamping machine replication (SMR) mode is at play. On the contrary, if both genomic and antigenomic strands are copied with the same efficiency, geometric replication (GR) takes place. 

The RM has important evolutionary and dynamical consequences in RNA viruses since it will involve different rates at which mutations accumulate thus affecting the statistical properties of the quasispecies \cite{Sardanyes2010}. Roughly, the distribution of mutations per genome within an infected cell for SMR is expected to be Poisson because mutants do not replicate \cite{Luria1951}. Consequently, the fraction of mutation-free genomes produced is given by the Poisson null class $e^{-\mu L}$, being $\mu$ the per-site mutation rate and $L$ the genome length. However, if all produced strands are used as templates, thus following GR, the distribution of mutant genomes conforms to the Luria-Delbr\"uck distribution \cite{Dewanji2005}. Here, the fraction of mutation-free genomes produced would depend on the number of replication rounds experienced, $\tau$, according to $e^{-\tau \mu L}$. Therefore, it is straightforward to see that GR will produce $f$ more mutant genomes than SMR according to the
equation $f = (1 - e^{-\tau \mu L})/(1 - e^{-\mu L})$. If only a fraction of the genomic strand progeny replicates, then the RM will be a mixture of SMR and GR that deviates from the Poisson expectation as much as the GR contribution.

Experimental data support different RMs for different viruses. For instance, bacteriophage T2 is thought
to replicate mostly by GR because the number of mutants per infected cell fails to fit a Poisson distribution \cite{Luria1951}. However,
phage $\phi$X174 data fit well the Poisson distribution, suggesting a SMR model \cite{Denhardt1966}. 
Within these two extremes, phage $\phi$6 slightly deviated from the Poisson expectation, an observation interpreted
as a result of a mixed model in which some progeny of positive strands was also able to replicate \cite{Chao2002}.

In this article I review recent advances in the dynamics of the RM for positive-sense single stranded RNA viruses. Then, previous results on a mathematical model describing the amplification dynamics of viral genomes with asymmetries due to different RMs are extended. Specifically, it is proved that the interior equilibrium of the phase plane is globally and asymptotically stable.
\begin{figure*}
\begin{center}
\includegraphics[width=.8 \textwidth]{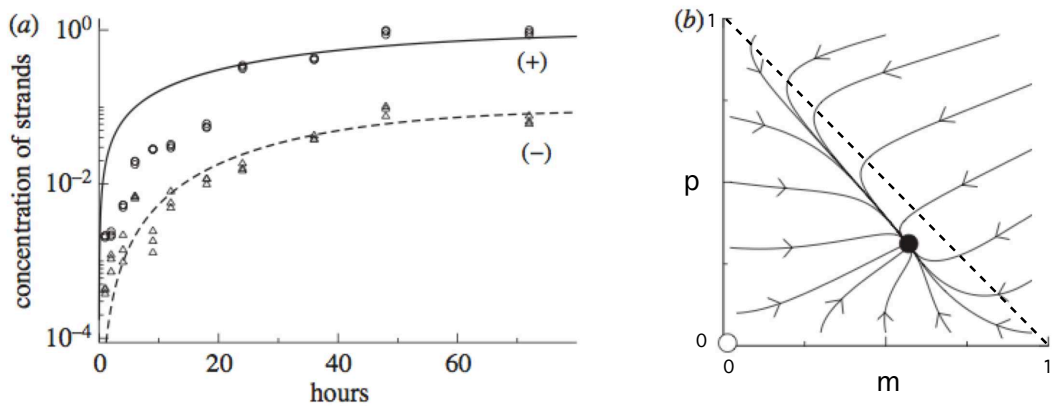}
\caption{(a) Dynamics of RNA genomes for Turnip mosaic virus. The dots and the triangles correspond to experimental data for positive and negative strands, respectively. The solid and the dashed lines are the simulated dynamics using the mathematical model (see \cite{Martinez2011} for further details). (b) Phase portrait for Eqs. \eqref{eq}: the black circle is the fixed point involving coexistence of viral genomes. The open circle is an unstable equilibrium. The dashed line is the boundary $p + m = K = 1$. The arrows indicate the direction of the flow.}
\end{center}
\end{figure*}

\section{Viral replication modes: recent advances}
The dynamics of viral RNA amplification was recently quantified for Turnip mosaic virus, and a simple mathematical model was used to fit the experimental data and infer the RM \cite{Martinez2011}. The same mathematical model was investigated considering natural infections and the fixed points and stability properties of the model were studied analytically and numerically \cite{Sardanyes2012}. The next Section extends the results presented in \cite{Sardanyes2012}.

\subsection{Dynamical evolution equations}
The mathematical model describing the within-cell amplification dynamics of both positive- and negative-sense strands investigated by Sardany\'es and co-workers \cite{Sardanyes2012} is given by the next system of differential equations, ${\bf{\dot{x}}} = f({\bf{x}})$, where ${\bf{x}} = (p, m)$ and:
\begin{equation}
 \frac{dm}{dt} = \alpha r p \left(1 - \frac{p+m}{K}\right) - \delta m,  \phantom{xxx}  \frac{dp}{dt} = r m \left(1 - \frac{p+m}{K}\right) - \delta p. \label{eq}
\end{equation}
The state variables $p$ and $m$ denote, respectively, the concentration of plus (positive) and minus (negative) viral RNA genomes. Parameter $r$ is the replication rate, $K$ corresponds to the cellular carrying capacity and $\delta$ is the genomes degradation rate. The variables span the two-dimensional phase space $\Gamma := \{(p, m) \in (\mathbb{R}^+)^2: 0 \leq p + m \leq K\}$.

Previous research on Eqs. \eqref{eq} computed local stability of the interior fixed point, given by $P^* = (p^*,m^*)$, which involves coexistence of both viral genomes polarities (see the phase portrait in Fig. 1(b) and \cite{Sardanyes2012}). Let us now extend the previous results by analyzing if such a fixed point is globally stable. We claim that $\Gamma$ is positively invariant. For that we analyze the direction of the vector field on the borders of $\Gamma$. On $(0,m)$ the vector field is $\dot{p} = rm (1 - m/K)$ and $\dot{m} = - \delta m$. On $(p,0)$ the vector field is $\dot{p} = - \delta p$ and $\dot{m} = \alpha r p (1 - p/K)$. Furthermore, on $(p,K-p)$ the vector field is $\dot{p} = - \delta p$ and $\dot{m} = - \delta (K-p)$. In the three cases the vector field points to the interior of $\Gamma$. Finally, we need to analyze the vector field on the corners of $\Gamma$ i.e., $(K,0)$ and $(0,K)$. On the point $(K,0)$, the vector field is  $\dot{p} = - \delta p$ and $\dot{m} = 0$. Since the second component of the vector field is $0$, we do not have enough information to state that solutions will enter into $\Gamma$. Then, we analyze how the initial condition $(K,0)$ will evolve in time. To do so we use a local analysis of this initial condition for positive time using a Taylor expansion. The corresponding solution is:
\begin{eqnarray} 
\varphi(t)  =   \varphi(0) + \dot{\varphi}(0)t + \frac{1}{2} \ddot{\varphi}(0) t^2 + ...  =  (K,0) + (-\delta K, 0) t + \frac{1}{2} \left(\delta^2 K, \alpha r \delta K \right) t^2 + ...,
\end{eqnarray}
since the border in $\Gamma$ is given by $K = 1$, it turns out that the coefficient of the second-order term of the Taylor expansion is: $$\frac{1}{2} (\delta^2 K, \alpha r \delta K)\lvert_{K=1} = \frac{1}{2} (\delta^2,\alpha r \delta); \phantom{x} \rm{with} \phantom{x}  \delta^2> 0, \phantom{x} \alpha r \delta > 0.$$ The previous calculations indicate that the solution with initial condition $(K, 0)$ will enter into $\Gamma$. Analogously, the same behavior is found for the point $(0,K)$ (results not shown). Hence, all previous calculations show that trajectories will enter into $\Gamma$ from the three borders of it, given by $(0,m)$, $(p,0)$, and $(p,K-p)$. By Poincar\'e-Bendixson theorem, the $\omega$-limit set of any initial condition on $\Gamma$ is not void and it is contained in $\Gamma$. By Dulac's criterion \cite{Strogatz2000} we will now prove that there is not a periodic orbit in $\Gamma$. For an autonomous planar vector field, Dulac's criterion states: let ${\bf{\dot{x}}} = f ({\bf{x}})$ be a continuously differentiable vector field defined on a simply connected subset $\Omega$ of the plane, i.e. $\Omega \subset \mathbb{R}^2$. If there exists a continuously differentiable, real-valued function $g({\bf{x}})$ such that $\nabla \cdot (g({\bf{x}}) f({\bf{x}}))$ has constant sign throughout $\Omega$, then there exist no closed orbits lying entirely in $\Omega$. Now we apply Dulac's criterion to our system, with $g = 1$, then:
\begin{equation}
\nonumber{
\nabla \cdot f({\bf{x}}) = \frac{\partial}{\partial p} \dot{p} + \frac{\partial}{\partial m} \dot{m}},
\end{equation}
is defined in our system as: $$\nabla \cdot f({\bf{x}}) =  \frac{\partial}{\partial p} \bigg(r m (1 - m - p) - \delta p\bigg) + \frac{\partial}{\partial m} \bigg(\alpha r p (1 - m - p) - \delta m \bigg) = - r (m + \alpha p) - 2 \delta.$$
Note that since $r, m, \alpha, p, \delta > 0$, the sign of $\nabla \cdot f({\bf{x}})$ is constant (i.e., $\nabla \cdot f({\bf{x}}) < 0)$ and since the domain $\Omega = \Gamma$ is simply connected and $g$ and $f$ satisfy the required smoothness conditions, we can conclude that there are no closed orbits in $\Gamma$. 

Since we have a unique fixed point in $\Gamma$, and we discarded the existence of a periodic orbit, according to the Poincar\'e-Bendixson then it must be the $\omega$-limit set of every initial condition and hence $P^*$ is globally asymptotically stable in $\Gamma$.

\section{Conclusions and prospectives}
The dynamics of mutation accumulation in viral pathogens is a key evolutionary parameter that still remains poorly understood. Such a subject has been addressed experimentally \cite{Luria1951,Denhardt1966, Chao2002} and both computationally and theoretically \cite{Sardanyes2010,Martinez2011,Sardanyes2012,Sardanyes2011}. Despite these previous investigations, several theoretical questions concerning the dynamics and the evolutionary consequences of different RMs still remain open. For instance: (i) what are the impacts of stochasticity in the dynamics of positive and negative strands in natural infections in terms of dynamical robustness? (ii) What are the expected deterministic dynamics in spatially-extended models of asymmetric replication? Do different RMs generate deterministically-driven spatial self-structuring? (iii) What is the interplay between epistasis and stochasticity or space in terms of dynamical robustness? Question (i) may be addressed by extending the model here presented to stochastic differential equations (e.g., Fokker-Planck equations). Question (ii) could be tackled with partial differential equations, and question (iii) should consider both previous theoretical approaches also incorporating mutant classes and different nonlinear interactions among mutations.  

\vspace{0.4cm}
{\bf{ACKNOWLEDGEMENTS}}. I especially thank Ernest Fontich for useful suggestions and Silvia Rubio for English corrections. I also thank Santiago F. Elena, Fernando Mart\'inez and Jose Antonio Dar\`os for sharing this research subject. This work was funded by the Bot\'in Foundation and by grant NSF PHY05-51164.


\end{document}